# Moiré-Tunable Localization of Simultaneous Type I and Type II Band Alignment in a MoSe$_2$/WS$_2$ Heterobilayer

Jiaxuan Guo[1], Zachary H. Withers[2,3], Ziling Li[4], Bowen Hou[1], Alexander Adler[2], Jianwei Ding[2], Victor Chang Lee[1], Roland K. Kawakami[4], Gerd Schönhense[5], Alice Kunin[6], Thomas K. Allison[2,3*], and Diana Y. Qiu[1*]

[1]Department of Materials Science, Yale University, New Haven, Connecticut 06511, USA
[2] Department of Physics and Astronomy, Stony Brook University, Stony Brook, New York 11794, USA
[3]Department of Chemistry, Stony Brook University, Stony Brook, New York 11794, USA
[4]Department of Physics, The Ohio State University, Columbus, Ohio 43210, USA
[5]Johannes Gutenberg-Universität, Institut für Physik, D-55099 Mainz, Germany
[6] Department of Chemistry, Princeton University, Princeton, New Jersey 08544, USA

**Abstract:**
Moiré heterobilayers exhibiting spatially varying band alignment and electron and hole localization that can be precisely controlled through the twist angle have emerged as exciting platforms for studying complex quantum phenomena. While most heterobilayers of transition metal dichalcogenides (TMDs) have a type II band alignment, the introduction of type I band alignment could enable stronger light-matter coupling and enhanced radiative emission. Here, we show through a combination of first-principles GW plus Bethe Salpeter equation (GW-BSE) calculations and time- and angle-resolved photoemission spectroscopy (tr-ARPES) measurements that contrary to previous understanding, the MoSe$_2$/WS$_2$ heterobilayer has a type I band alignment at large twist angles and simultaneous regions of type I and type II band alignment due to the structural reconstruction in different high symmetry regions at small twist angles. In tr-ARPES, consistent with our calculations, a long-lived electron population is only observed in MoSe$_2$ for samples with large twist angles, while in samples with small twist angles, signals from two distinct long-lived excitons are observed. Moreover, despite the near degeneracy of the conduction bands of the two layers, no excitonic hybridization occurs, suggesting that previously observed absorption peaks in this material arise from lattice reconstruction. Our findings clarify the complex energy landscape in MoSe$_2$/WS$_2$ heterostructures, where the coexistence of type I and type II band alignment opens the door to moiré-tunable optoelectronic devices with intrinsic lateral heterojunctions.

* diana.qiu@yale.edu
* thomas.allison@stonybrook.edu

*Introduction*

Two-dimensional transition metal dichalcogenides (TMDs) exhibit remarkable electronic and optical properties, making them exciting platforms for exploring excited-state physics. In monolayer TMDs, the combination of quantum confinement and reduced dielectric screening enhances Coulomb interactions, leading to the existence of strongly-bound excitons, with binding energies of around 0.5 eV, as well as other multiparticle excitations [1–7]. These excitons dominate the linear and nonlinear [8] optical properties of monolayer TMDs and facilitate strong light-matter interactions that enable the concurrent manipulation of charge, spin, and valley degrees of freedom [9–13].

When two monolayer TMDs, or other layered 2D materials, with a small lattice mismatch or twist angle are vertically stacked, a moiré pattern that can span several nanometers emerges. This moiré pattern creates a long-range spatially varying potential that can be leveraged to engineer the electronic band structure and explore correlated electron states, such as Mott-insulators [14–19], Wigner crystals [14,15,17,20–22] and various topological states [23–34]. Unlike in twisted bilayer graphene [35,36], at small twist angles, TMDs undergo substantial lattice relaxation leading to an atomic reconstruction with strained regions where the stacking is nearly commensurate [37–43]. These strained regions host a large variety of tunable, localized exciton states, including interlayer excitons [44], inter- and intralayer moiré excitons [45,46], where the electron and hole sit at different moiré lattice sites, and hybrid states, where inter- and intralayer excitons hybridize [47]. Moreover, the moiré-induced localization enables access to excitons with a distribution of finite momenta, whose spatial confinement has the potential to be exploited as quantum dots and single-photon emitters [48].

While heterobilayers of $MX_2$ TMDs (M = Mo, W; X = S, Se) mostly exhibit type II band alignment, in which the conduction band minimum (CBM) and valence band maximum (VBM) lie in different layers [49–51], the band alignment of $MoSe_2/WS_2$ heterobilayer remains the subject of considerable debate due to the near-degeneracy of conduction bands within the two layers. Early first-principles density functional theory (DFT) and many-body GW calculations indicated that $MoSe_2/WS_2$ exhibits a type II band alignment, with the valence band in the $MoSe_2$ layer and the conduction band in the $WS_2$ layer [49–51]. Assumptions about the type II band alignment combined with the apparent near degeneracy of the conduction band in both layers also lead to the hypothesis that three distinct low energy peaks in the optical absorption spectrum arise from the hybridization of interlayer and intralayer exciton states [52]. However, more recent experiments and corresponding model calculations indicate that $MoSe_2/WS_2$ actually has a type I band alignment with both the VBM and CBM in the MoSe$_2$ layer. Experimentally, angle-resolved photoemission spectroscopy (ARPES) measurements incorporating alkali doping to populate the conduction band show some evidence for the CBM being in MoSe$_2$ [53], and reflectance contrast spectra as a function of doping density reveal that charged excitons only form in the MoSe$_2$ layer in large twist-angle MoSe$_2$/WS$_2$ heterostructures, consistent with a type I band alignment [54,55]. Theoretically, a phenomenological model for intra- and interlayer excitons, fit to the experimental differential reflectance signal of MoSe$_2$/WS$_2$ as a function of the out-of-plane electric field, suggests that the MoSe$_2$ CBM is lower than the WS$_2$ CBM by about 50 meV [56]. The hybrid exciton hypothesis [47,52] for the splitting of exciton peaks near the MoSe$_2$ A resonance has also been challenged, as reflectance spectroscopy shows no evidence of out-of-plane dipoles near the MoSe$_2$ A peak [54,56].

In this work, we employ a combination of first-principles GW and GW plus Bethe Salpeter equation (GW-BSE) calculations and time- and angle-resolved photoemission spectroscopy (tr-ARPES) to investigate the quasiparticle (QP) band alignment of MoSe$_2$/WS$_2$ heterobilayer and the character of the low-energy excitons near the MoSe$_2$ A resonance. We find that in the absence of atomic reconstruction, consistent with large twist angles, the MoSe$_2$/WS$_2$ heterobilayer exhibits a uniform type I band alignment after many-electron interaction effects are included at the GW level. Consistent with other TMD bilayers [41,42], at small twist angles, there is a large atomic reconstruction resulting in strained high symmetry stacking regions, where the strain in each layer considerably alters the electronic structure. The large reconstruction results in the simultaneous emergence of alternating regions of type I and type II band alignment in the moiré supercell of MoSe$_2$/WS$_2$. In tr-ARPES measurements of excitons, we observe long-lived electron population only in MoSe$_2$ for samples with large twist angle, while for samples with small twist angles and large moiré unit cells we observe both long-lived intra- and interlayer exciton species with different binding energies, consistent with the co-existence of type I and type II band alignment in our calculations. Finally, our GW-BSE calculations on moiré-free heterobilayers that are strained to match the moiré-driven reconstruction reveal that contrary to previous hypotheses of the existence of hybrid excitons [47,52], the lowest-energy excitons in the MoSe$_2$/WS$_2$ heterostructure are purely intralayer excitons, with negligible interlayer character. We note that the strain induced by the lattice reconstruction shifts absorption peaks in different high-symmetry stacking regions, and we hypothesize that the co-existence of the three distinct stacking regions contributes to the formation of the three low-energy exciton peaks seen in the optical spectrum at small twist angles [47,54,56] and the two distinct lowest energy excitons in our tr-ARPES measurements.

*Computational Methods*

We start by performing DFT calculations as implemented in the Quantum ESPRESSO code [57] with the Perdew-Burke-Ernzerhof (PBE) generalized gradient approximation (GGA) [58] for the exchange-correlation energy. Then, we calculate the QP band alignment and exciton states within the GW and GW-BSE formalism on top of the DFT mean field. The GW and GW-BSE calculations, with a fully relativistic spinor formalism, are performed with the BerkeleyGW package [59–62]. We construct twisted bilayer TMDs with the help of the TWISTER code [63], relax atomic structures using the Stillinger−Weber (SW) force field [64] and the parametrized Kolmogorov−Crespi potential [65–67] as implemented in the LAMMPS package [68], and visualize atomic strain using OVITO [69] and VESTA [70,71]. Additional computational details can be found in the Supplemental Material.

*Band Alignment at Different Twist Angles*

We start by calculating the absolute band alignment of individual isolated monolayers with respect to the vacuum level [49–51]. Fig.1 (a) and (d) show the vacuum-aligned DFT band structure of the individual monolayers at the PBE level, which exhibits type II band alignment with the valence band in the $MoSe_2$ layer and the conduction band in the $WS_2$ layer. Spin-orbit coupling splits the conduction band of $WS_2$ by 30 meV, and the CBM of $MoSe_2$ is nearly degenerate with $WS_2$, lying only 27 meV above the $WS_2$ CBM, between the spin-orbit split bands, consistent with previous DFT calculations [50]. However, when we consider many-electron interactions and correct the band energies by the GW approximation, the $MoSe_2$ CBM becomes lower than the $WS_2$ CBM by 73 meV resulting in a type I band alignment (Fig.1(b) and (e)). Of course, in a real heterostructure, even if wavefunctions of the individual monolayers do not hybridize at the band edge, the electronic states are still screened by both layers. To isolate the role of bilayer screening on the band alignment, we then perform GW calculations for each monolayer using screening that includes the polarizability of both layers in the random phase approximation (RPA) [72,73], essentially treating the second layer as a substrate whose only contribution is to the polarizability [3,74–76]. The bilayer screening renormalizes the QP bandgap, reducing the gap in $WS_2$ from 2.48 eV to 2.22 eV and the gap in $MoSe_2$ from 2.15 eV to 1.99 eV, but the overall type I GW band alignment remains unchanged (Fig. 1 (c) and (f)).

In a real heterobilayer, the band alignment may also be influenced by atomic reconstruction [42,77–80]. Next, we explore the effect of atomic reconstruction on the band alignment at both large twist angles, where the bandstructure at the GW level can be explicitly calculated, and small twist angles, where we approximate large area reconstructions as regions of nearly uniform stacking (see the Section IV in Supplemental Material). We start by exploring the effect of twist angle on atomic reconstruction by constructing large supercells with a $MoSe_2$ monolayer stacked on a WS$_2$ monolayer at 8 different twist angles (1.8º, 11.2º, 19.1º, 27.8º, 35.5º, 43.9º, 51.7º, 60º). Then, we relax the supercells using molecular dynamics (MD) following previous workflows [42]. The size of the moiré superlattice and the maximum strain due to reconstruction in each layer after relaxation are reported in Table 1. The smallest moiré supercell appears at 43.9° with only 75 atoms and a moiré lattice constant of 1.1 nm. In the small moiré cell, the change in lattice constants due to atomic strain is ±0.006 Å, indicating that there is almost no reconstruction. This small amount of strain has a negligible effect on the band energies (see Fig. S2 in Supplemental Material). When the twist angle is near 0° or 60°, the moiré supercell is much larger, extending 6-7 nm, and includes thousands of atoms. At these twist angles, there is a large reconstruction. For example, in the case of a 60° twist angle, the lattice constant of the both the MoSe$_2$ and WS$_2$ layers experience a maximum strain of 0.9%. At this level of strain, band energies shift by hundreds of milli-electronvolts (see Fig. S2 in the Supplemental Material), which is considerably larger than the CBM offset between MoSe$_2$ and WS$_2$ in the absence of moiré reconstruction, suggesting the possibility of a twist-angle dependent change in band alignment.

We start by analyzing the small moiré supercell case, for a twist angle of 43.9° where the reconstruction is small. Fig. 2 (a) and (b) show the crystal structure and corresponding Brillouin Zone (BZ). DFT and GW calculations are performed directly on the relaxed supercell. Fig. 2 (c) shows the DFT band structure colored according to the projected density of states on each layer. The band edges of the MoSe$_2$ and WS$_2$ layers are at the $\gamma$ and $\kappa$ points of the moiré BZ, respectively, as expected from the unfolding of the moiré BZ shown in Fig. 2 (b). At the DFT level, MoSe$_2$/WS$_2$ has type II band alignment, which is identical to that of the isolated monolayers. At the GW level (Fig. 2 (d)), the MoSe$_2$ CBM lies 106 meV below the WS$_2$ CBM, so the heterostructure has type I band alignment, which is again consistent with calculations on the isolated monolayers. Therefore, at 43.9°, or more generally, at large twist angles when the moiré lattice is small, the moiré potential does not change the band alignment of MoSe$_2$/WS$_2$, and the heterostructure has a type I band alignment across the entire moiré lattice.

We then investigate the case where the twist angle is near 60°, and the moiré superlattice is large. Fig. 3(a) shows the moiré superlattice. Three distinct stacking regions that form after the relaxation are labeled $H_X^M$, denoting a configuration where the central atom is a metal atom stacked on a chalcogen atom in the neighboring layer, $H_M^M$, denoting a central metal atom stacked on a metal atom, and $H_X^X$, denoting a central chalcogen in one layer stacked with a chalcogen in the other layer. Fig. 3(b) shows the strain distribution in the same moiré supercell. Positive (negative) percentages mean that the lattice is under compressive (tensile) strain. We see that the MoSe$_2$ layer is under tensile strain in the $H_X^X$ region and compressive strain in the $H_X^M$ region, while WS$_2$ experiences the opposite strain. In the $H_M^M$ region both materials are relatively unstrained. Within a radius of 1.2 nm, which roughly corresponds to the radius of the exciton, the standard deviation of the lattice constants in $H_X^M$, $H_X^X$ and $H_M^M$ are 0.002 Å, 0.005 Å, and 0.006 Å, respectively (see the section IV in Supplemental Material).

Since the radius of each stacking region for the large moiré superlattice is comparable to the exciton radius [2], and the strain is relatively uniform within each region, it is reasonable to approximate each region of the moiré superlattice as a heterobilayer with a quasi-uniform strain. We start by determining the change in the band alignment due to the strain in each stacking region at the GW level. Since each stacking is not perfectly commensurate, it is not possible to construct a unit cell with the correct average strain. Instead, to obtain the correct band alignment, we start by calculating the absolute VBM and CBM energies with respect to the vacuum level of monolayer MoSe$_2$ and WS$_2$ and incorporate screening taken from the bilayer with the correct stacking configuration. Fig. 3(c) shows the CBM and VBM of MoSe$_2$ and WS$_2$ with lattice constants under different strains. Circles indicate unstrained lattice constants in the $H_M^M$ region. Squares mark the average lattice constant in the $H_X^M$ and $H_X^X$ regions. Triangles mark maximum or minimum lattice constants appearing at the boundary of $H_X^M$ and $H_X^X$ regions. Our results are consistent with the widely recognized fact that tensile strain decreases both the band gap and the absolute energy of the band edges at the K point in TMDs, whereas compressive strain increases them [81,82]. In the $H_X^X$ region, MoSe$_2$ is stretched while WS$_2$ is compressed, so the MoSe$_2$ CBM decreases and the WS$_2$ CBM increases, resulting in type I band alignment with a band gap of 1.92 eV compared to the bandgap of 1.99 eV found in the unstrained bilayer. The opposite trend is observed in the $H_X^M$ region, leading to type II band alignment with a bandgap of 1.86 eV.

Our GW calculations show that MoSe$_2$/WS$_2$ has type I band alignment at large twist angles, where the moiré lattice and reconstruction are small, and both type I and II band alignment in different high-symmetry regions at small twist angles, where the moiré lattice and reconstruction are large. The coexistence of two types of band alignment introduces the possibility of hosting both intralayer and interlayer excitons in different regions of space within the same heterobilayer.

*Excitons and Absorption Spectra*
Next, we investigate excitons that can arise in these different regions within the *ab initio* GW plus Bethe Salpeter equation (GW-BSE) approach [60,62]. To avoid the computational demand of performing calculations on the full moiré supercell, we again make the reasonable approximation that each stacking region is well-described as a uniform heterobilayer. To account for both interlayer and intralayer excitons, we construct a strained commensurate unit cell of MoSe$_2$/WS$_2$.

Because the commensurate cell has slightly different strain than the high symmetry stacking regions in the reconstructed moiré superlattice, we then treat the change in energy due to strain from the moiré lattice as a first order perturbation on the energies of the commensurate bilayer. This is reasonable since the strain is small and the wavefunctions do not hybridize near the K point.

Fig.4 (a), (c), and (e) show the calculated absorption spectrum corresponding to each high-symmetry stacking region, and (b), (d), and (f) show the contribution of different QP bands projected onto the MoSe2 or WS2 layer to the exciton wavefunction at each energy. In the Tamm-Dancoff approximation [62,83], the exciton can be written as a linear combination of electron-hole pairs $|S\rangle = \Sigma_{vck} A^S_{vck} |vc\mathbf{k}\rangle$, where $A^S_{vck}$ denotes the electron-hole amplitude at the $\mathbf{k}$ point in the reciprocal space. $v$ and $c$ are indices for the valence and conduction bands respectively and $S$ is the principle quantum number of the exciton. In Fig 4 (b), (d), and (f), every exciton is depicted as a column of points, where the size of each dot corresponds to the square of the electron-hole amplitude, $\sum_{ck} |A^S_{vck}|^2$ for valence-band states, and $\sum_{vk} |A^S_{vck}|^2$ for conduction-band states.

From this decomposition, the lowest energy exciton appears in the $H^X_X$ region where the bandgap is smallest. It is a purely intralayer exciton corresponding to the MoSe2 A exciton. We note that contrary to previous hypotheses [47,52], the MoSe2 A exciton does not include any interlayer hybridization and comes purely from the MoSe2 layer. In our calculations, the GW correction lifts the near degeneracy of the MoSe2 and WS2 conduction band edges, which was the basis of previous hypotheses of inter- and intralayer hybridization [52]. However, even when the two conduction bands are nearly degenerate, we still do not see hybrid excitons in our BSE calculation. Some higher energy interlayer excitons do exhibit a degree of hybridization due to interaction with the continuum of lower energy intralayer excitons.

Next, we turn to understanding the absorption features in different stacking regions. The absorption spectrum of MoSe2/WS2 with a near 60-degree twist is dominated by three distinct exciton peaks at 1.39 eV, 1.46 eV, and 1.49 eV. As previously noted, the $H^X_X$ region is dominated by a peak corresponding to the intralayer MoSe2 A exciton at 1.39 eV, which has a binding energy of 0.53 eV. For convenience, we will refer to this as the $A^X_X$ exciton. In the $H^M_M$ region, the band alignment remains type I, and the lowest energy exciton is still the intralayer MoSe2 A exciton ($A^M_M$), now shifted to a higher energy of 1.46 eV. It has a similar binding energy of 0.53 eV. In the $H^M_X$ region, since there is type II band alignment, the lowest energy exciton is an interlayer exciton ($I^M_X$) with the hole in the MoSe2 layer and the electron in the WS2 layer. The interlayer exciton is nearly dark, and the absorption spectrum is dominated by the bright intralayer exciton in MoSe2 ($A^M_X$) at 1.49 eV. In the $H_X^M$ region, the binding energy of the interlayer exciton is 0.43 eV, while the binding energy of the intralayer exciton is 0.53 eV. Although it is hard to observe the interlayer exciton directly in reflectance contrast experiment because of its weak oscillator strength, it can be brightened by applying out-of-plane electric field. The existence of the interlayer exciton in the $H_X^M$ region at an energy between $A_X^M$ and $A_X^X$ is consistent with experimental observation of a weak dispersion of the second and third absorption peaks under a perpendicular electric field [54,56]. The energy difference between the $A_X^M$ and $A_X^X$ peaks is about 100 meV, which is similar to the difference in energy between the first and third absorption peaks observed in previous optical measurements [54,56]. Thus, we attribute the three low-energy peaks in the experimental spectrum to intralayer excitons in the three different high-symmetry regions. While this calculation does not account for the possibility of intralayer charge transfer excitons [78], it sufficiently captures interlayer and intralayer excitons localized at each moiré site, which are expected to be much brighter in the absorption spectrum than charge transfer excitons.

*TR-ARPES*

To experimentally determine the nature of the excitons produced in different heterostructures and also compare dark excitons, we apply tr-ARPES to $MoSe_2/WS_2$ samples with different twist angles and different layer stacking orders. For samples with large twist angles, the K valleys of the two layers are easily discernible in momentum-space maps and tr-ARPES directly reports on the layer occupation of the electron component of the exciton wave function after photoexcitation. For samples with small twist angles, where the K valleys of the two layers coincide in momentum space, tr-ARPES can still delineate between inter- and intralayer excitons via the extreme surface sensitivity of the technique. Photoelectron signals from the lower layer are heavily attenuated due to the very small electron mean free path at ~20 eV energy [53,84,85], such that our measurement predominantly measures the electron population in the top-most layer of the heterostructure. The Stony Brook time-resolved ARPES beamline and its application to 2D materials have been described previously [86–89]. Heterostructure regions of the samples are selected using an aperture in a real-space image plane of the momentum microscope (see Supplemental Material and [90]). The high data rate of the Stony Brook beamline enables us to perform a systematic series of measurements on many samples. In all experiments, the samples are excited with 2.4 eV pump photon energy, which initially populates many exciton states in both layers. In this study, we focus on the long-lived excitons persisting after initial relaxation.

Fig. 5 (a-f) show 2D cuts of the 3D ARPES data along the K-Γ-K direction for 6 different samples, for pump-probe delays greater than 5 ps, recorded with 25.2 eV photon energy. Fig. 5 (a) and 5 (b) show signals from monolayer $MoSe_2$ and $WS_2$, Fig. 5 (c) and 5 (d) are for heterostructures with twist angles near 60º (the small twist/large moiré supercell condition) and Fig. 5 (e) and 5 (f) are for heterostructures near 40º (the large twist/small moiré supercell condition). For the monolayers, the energy scale is referenced to the global VBM at K. For the bilayers, the maximum at K is not clearly visible due to the photoemission matrix element effect, therefore we reference energies to the local VBM at Γ ($E_\Gamma$). The monolayer signals show the VBM at K, while the heterobilayer signals show that the VBM at Γ is nearly degenerate with the MoSe$_2$ VBM at K, consistent with our calculations and previous work [91]. Similar to previous work and consistent with our GW-BSE calculations, in the bilayers we observe a ~500 meV splitting of the valence bands at Γ due to strong hybridization of the chalcogen p$_z$ orbitals.

Exciton signals are observed above the VBM but below the CBM due to the exciton binding energy [92–97]. Fig. 5 (g) and 5 (h) show the pump-probe delay dependence of the exciton signals for the different samples. Monolayer exciton signals (shown in both Fig. 5 (g) and 5 (f)) appear at energies consistent with PL measurements [98–100], and decay on the 10-100 ps timescale, consistent with our previous work [88]. For the bilayers, signals recorded from MoSe$_2$ on WS$_2$, which are sensitive to the electron population in the MoSe$_2$ layer, show the same dynamics as the intralayer excitons in monolayer MoSe$_2$ regardless of the twist angle. The exciton signal appears at 1.59±0.03 eV above the $E_\Gamma$, which is in reasonable agreement with the calculated excitation energy of the $A_X^X$ exciton of 1.39 eV. In contrast, signals recorded from WS$_2$ on MoSe$_2$, which are sensitive to the electron population in the WS$_2$ layer, show a strong twist-angle dependence, with no long-lived signal in the large-twist-angle samples. This suggests that the excitons in large-twist-angle samples are intralayer excitons with both the hole and the electron residing in the MoSe$_2$ layer, consistent with the type I band alignment predicted in our calculations. For samples with near 60°. twist angle, we observe a long-lived electron population in the WS$_2$ layer, appearing at 111±13 (statistical error) ±42 (systematic error) meV higher photoelectron energies than signals from the MoSe$_2$ population. This energy shift is consistent with the expected signature of an interlayer exciton $I_X^M$ in the $H_X^M$ stacking region. The calculated excitation energy of $I_X^M$ is 40 meV higher than the energy of the $A_X^X$ exciton. In addition, the valence band edge in the $H_X^M$ region is 47 meV higher than the $H_X^X$ region, which means that photoionization of the $A_X^X$ exciton leaves the hole unrelaxed by 47 meV with a corresponding reduction in the photoelectron energy referenced to the global VBM, leading to an overall energy shift of 87 meV for the tr-ARPES signal of $I_X^M$ compared to $A_X^X$.

*Conclusion*

In summary, we have performed first-principles GW calculations on a MoSe$_2$/WS$_2$ heterobilayer and show that the heterobilayer has a type I band alignment when the twist angle is large and the moiré length is small. At small twist angles, when the moiré length is large, we find that the large atomic reconstruction introduces significant strain, resulting in regions of both type I and type II band alignments at different moiré lattice sites. This spatially varying type I and type II band alignments could be an interesting platform for exploring novel phases of electrons and charged excitons [101]. Moreover, the system would form a natural lateral heterojunction with the energy landscape directing electron flow to the lower energy type I region, giving rise to the possibility of enhanced emission and population inversion. Our BSE calculations show that the absorption spectrum is dominated by three low-energy exciton peaks corresponding to the $MoSe_2$ A exciton in different stacking regions, with energy splitting in good agreement with experiment [54,56]. We observe no low-energy hybrid excitons, and the absorption peaks appear to be adequately explained by the lattice reconstruction. Our tr-ARPES measurements, which see a long-lived exciton population only in the $MoSe_2$ layer at large twist angles and at the energy of the intralayer exciton in $MoSe_2$ and the interlayer exciton in $WS2$ at small twist angles, is consistent with a transition from pure type I band alignment at large angles to the coexistence of type I and type II at small twist angles. Our work provides a comprehensive picture of the complex electronic and excitonic landscape in $MoSe_2/WS_2$ heterobilayers, where the band alignment and optical properties can be precisely tuned through twist angles. The discovery of coexisting type I and type II band alignments within a single supercell opens pathways for moiré-engineered optoelectronic devices with intrinsic lateral heterojunctions and tunable light-matter interactions.

*Acknowledgements*

This work was primarily supported by the U.S. Department of Energy, Office of Science, Basic Energy Sciences under Early Career Award No. DE-SC0021965, which supported theory and first principles calculations on moiré heterostructures. Development of the BerkeleyGW code was supported by Center for Computational Study of Excited-State Phenomena in Energy Materials (C2SEPEM) at the Lawrence Berkeley National Laboratory, funded by the U.S. Department of Energy, Office of Science, Basic Energy Sciences, Materials Sciences and Engineering Division, under Contract No. DE-C02-05CH11231. The calculations used resources of the National Energy Research Scientific Computing (NERSC), a DOE Office of Science User Facility operated under contract no. DE-AC02-05CH11231, and the Texas Advanced Computing Center (TACC) at The University of Texas at Austin. TR-ARPES work was supported by the U.S. Department of Energy, Office of Science, Office of Basic Energy Sciences under award number DE-SC0022004 and the Air Force Office of Scientific Research under FA9550-20-1-0259. Z.H.W. acknowledges support from the U.S. National Science Foundation Graduate Research Fellowship Program. Fabrication of the samples (Z.L. and R.K.K.) was supported by the U.S. Department of Energy, Office of Science, Basic Energy Sciences under Award No. DE-SC0016379 and the AFOSR/MURI project 2DMagic under Award No. FA9550-19-1-0390.

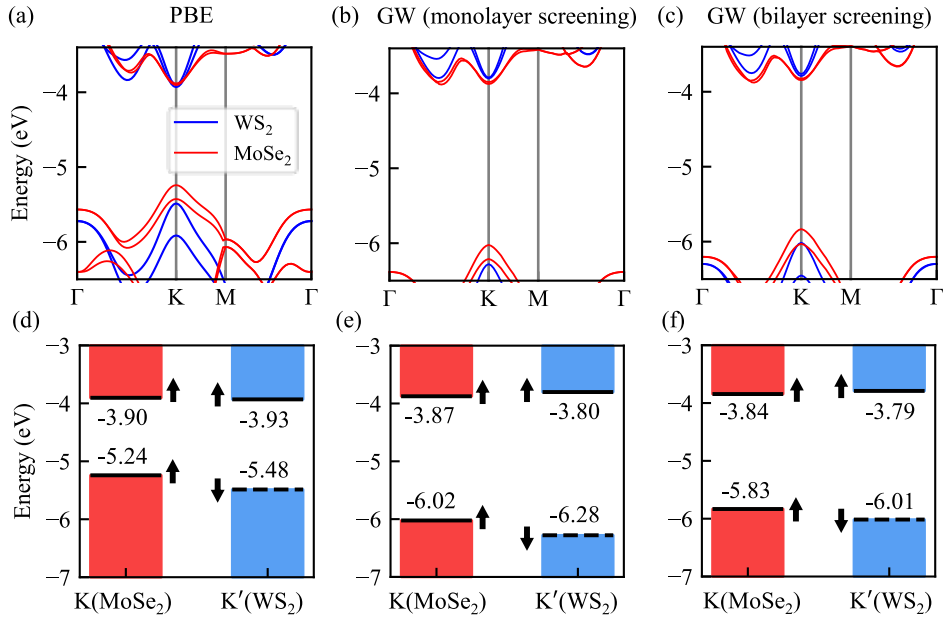

Fig. 1. (a)-(c) Band structures of monolayer MoSe$_2$ (red) and WS$_2$ (blue) calculated by DFT, GW with monolayer and bilayer dielectric screening. Calculations are performed individually for each monolayer, and the bands are shifted so that the vacuum level is set to zero to obtain the band alignment. The bilayer dielectric screening is obtained from the heterobilayer MoSe$_2$/WS$_2$. (d)-(f) Corresponding band alignments of MoSe$_2$/WS$_2$ at the K/K' point. Solid (dashed) lines indicate spin up (down) bands. DFT calculations show a type II band alignment, while one-shot G$_0$W$_0$ calculations (both monolayer and bilayer screening) indicate type I band alignment.

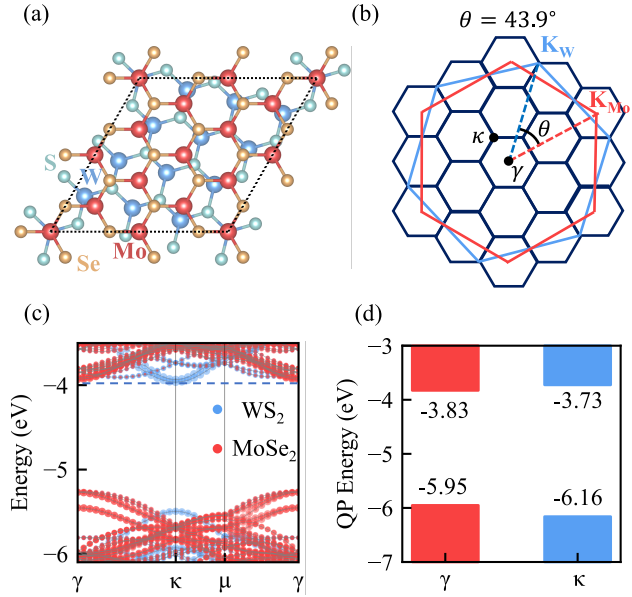

Fig. 2. (a) Atomic structure of 43.9° twist-angle MoSe$_2$/WS$_2$ (top view). (b) Brillouin Zones (BZ) of MoSe$_2$(red), WS$_2$(blue), and moiré heterostructure (black). The MoSe$_2$ K point (K$_{Mo}$) is folded to the $\gamma$ point of the moiré BZ, while the WS$_2$ K point (K$_w$) is folded to the $\kappa$ point of the moiré BZ. (c) DFT band structure colored according to the projected density of states on each layer. The blue dashed line marks the position of the WS$_2$ CBM, which is lower than the MoSe$_2$ CBM by 34 meV, highlighting the type II band alignment at the DFT level. (d) Quasiparticle (QP) band edge at $\gamma$ and $\kappa$ of the moiré BZ obtained from GW calculations. The MoSe$_2$ CBM (red), located at $\gamma$, is lower than WS$_2$ CBM (blue), located at $\kappa$, by 102 meV. At the GW level, MoSe$_2$/WS$_2$ has a type I band alignment.

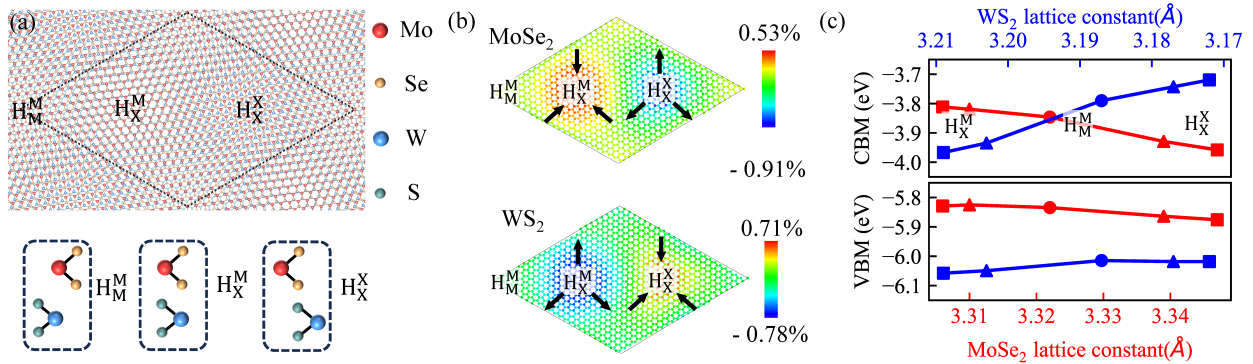

Fig. 3. (a) Moiré lattice for 60° twist angle MoSe$_2$/WS$_2$ (top). H$_X^M$, H$_X^X$, and H$_M^M$ (where M is the transition metal atom and X is the chalcogen atom) are three high-symmetry regions in the unit cell. Their stacking configurations are shown in the black dashed box (bottom). (b) Strain distribution in the same moiré lattice plotted as a variation in local lattice constants from equilibrium. Red (positive) indicates compressive strain; blue (negative) indicates tensile strain. In the $H_X^M$ region, MoSe$_2$ is under compressive strain while WS$_2$ is under tensile strain. The $H_X^X$ region shows the opposite. (c) GW band edges of strained monolayer MoSe$_2$ and WS$_2$ with bilayer screening. All energies are aligned relative to the vacuum level. In each region, WS$_2$ and MoSe$_2$ are strained in opposite directions, so MoSe$_2$ is plotted with increasing strain, and WS$_2$ is plotted with decreasing strain. Circles are unstrained lattice constant in $H_M^M$. Squares are average length of the lattice constants in high-symmetry regions $H_X^M$ and $H_X^X$. Triangles are maximum or minimum length of the lattice constant, appearing at the boundary of regions $H_X^M$ and $H_X^X$. Results indicate type I band alignment in the $H_X^X$ region and type II band alignment in the $H_X^M$ region.

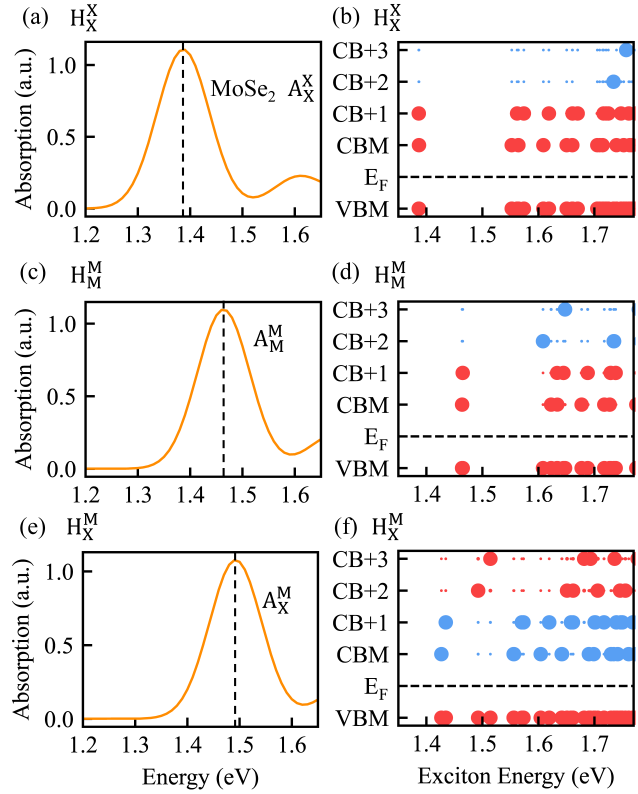

Fig. 4 (a), (c), and (e). Absorption spectra of MoSe$_2$/WS$_2$ in different high-symmetry regions with a constant broadening of 0.05 eV. BSE calculations are performed in a patch of 0.193 Å$^{-1}$ around the K point in the BZ. The highest peak corresponds to the MoSe$_2$ A exciton and its center is marked with a black dashed line. The exciton peak position shifts due to different strain in the three high-symmetry regions. (b), (d), and (f). Contributions of different QP bands to each exciton state. MoSe$_2$ valence and conduction band contributions are in red, while WS$_2$ conduction band contributions are in blue. The exciton state can be written as a linear combination of electron-hole pairs $|S\rangle = \Sigma_{vck} A_{vck}^S |vck\rangle$, where $A_{vck}^S$ is the electron-hole amplitude for the valence band $v$ and conduction band $c$ at k-point $\bm{k}$. The size of each dot is proportional to $\sum_{ck} |A_{vck}^S|^2$ for valence-band states, and $\sum_{vk} |A^S_{vck}|^2$ for conduction-band states. Intralayer and interlayer excitons are clearly distinguishable, indicating that no hybrid excitons contribute to the main peak in the absorption spectrum in each stacking region.

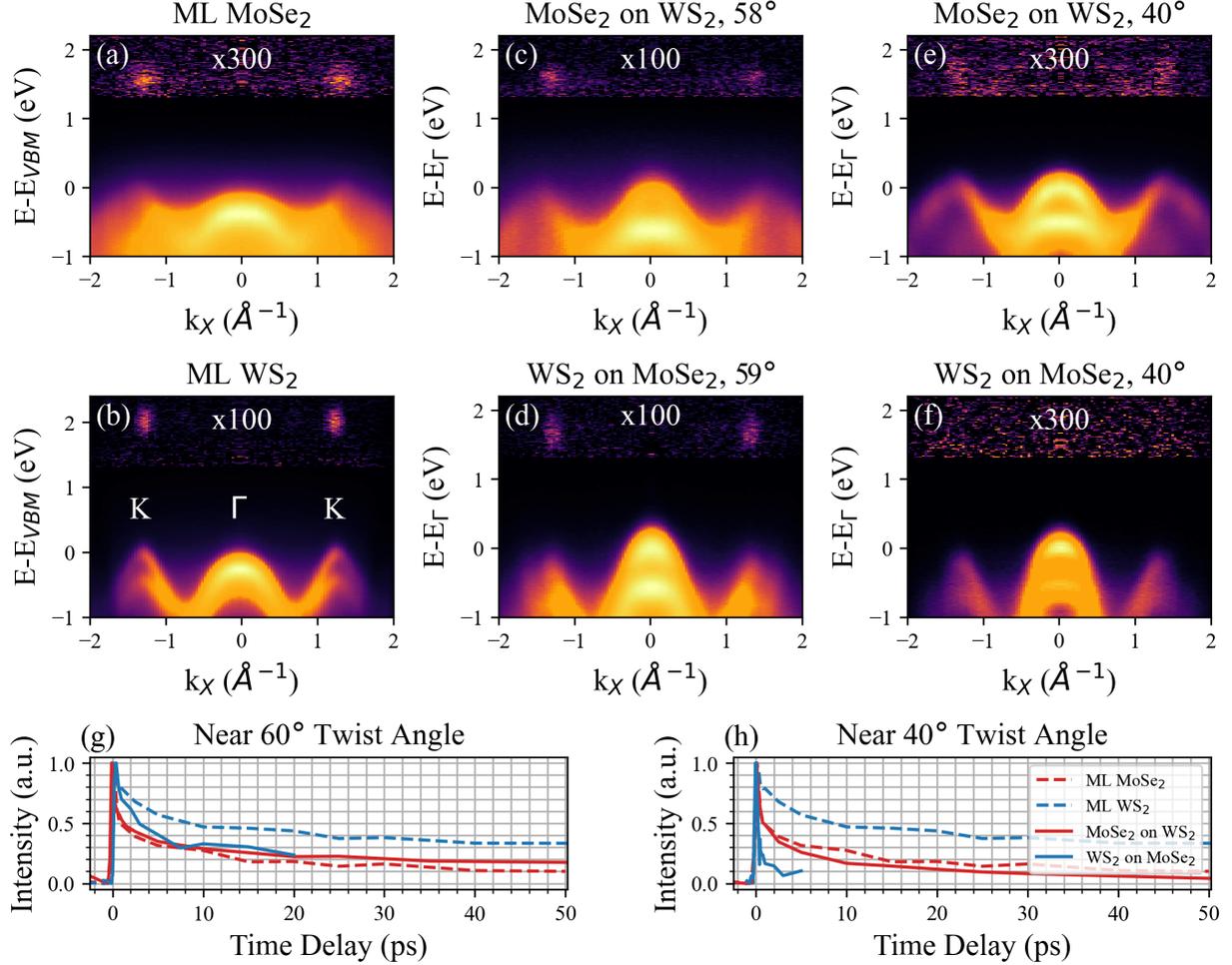

Fig. 5 tr-ARPES of ML-WS$_2$, ML-MoSe$_2$, and twisted heterobilayers. (a)-(f) K-Γ-K slice constructed from tr-ARPES data set by summing all pump/probe delays 5 ps and longer. The signal above 1.3 eV is multiplied by the constant displayed to show the exciton signal within the band gap. A nonlinear color scale is used to make all features visible. Logarithmic and linear scale versions appear in the Supplemental Materials. (g-h) Normalized intensity traces comparing the integrated exciton decay of the monolayer samples to heterobilayers with (g) small and (h) large relative twist angles.

Table 1. MD forcefield relaxation of MoSe$_2$/WS$_2$ moiré supercells at different twist angles. For the relaxed structure at each twist angle, the moiré length, the number of atoms in the supercell, and the maximum strain amplitude in the relaxed WS$_2$ layer and MoSe$_2$ layer are shown.

| Angle | Moiré Length | Number of Atoms | Max. Strain (WS$_2$) | Max. Strain (MoSe$_2$) |
| --- | --- | --- | --- | --- |
| 1.8º | 6.1 nm | 2112 | 0.030 Å (1.2%) | 0.027 Å (0.8%) |
| 11.2º | 2.8 nm | 456 | 0.012 Å (0.4%) | 0.012 Å (0.4%) |
| 19.1º | 3.5 nm | 699 | 0.007 Å (0.3%) | 0.008 Å (0.3%) |
| 27.8º | 5.8 nm | 1884 | 0.005 Å (0.2%) | 0.005 Å (0.2%) |
| 35.5º | 3.3 nm | 627 | 0.005 Å (0.2%) | 0.006 Å (0.2%) |
| 43.9º | 1.1 nm | 75 | 0.006 Å (0.2%) | 0.006 Å (0.2%) |
| 51.7º | 6.5 nm | 2373 | 0.011 Å (0.3%) | 0.013 Å (0.4%) |
| 60.0º | 7.5 nm | 3462 | 0.025 Å (0.8%) | 0.030 Å (0.9%) |

# Supplemental Material

Jiaxuan Guo[1], Zachary H. Withers[2,3], Ziling Li[4], Bowen Hou[1], Alexander Adler[2], Jianwei Ding[2], Victor Chang Lee[1], Roland K. Kawakami[4], Gerd Schönhense[5], Alice Kunin[6], Thomas K. Allison[2,3*], and Diana Y. Qiu[1*]

[1]Department of Materials Science, Yale University, New Haven, Connecticut 06511, USA
[2] Department of Physics and Astronomy, Stony Brook University, Stony Brook, New York 11790, USA
[3]Department of Chemistry, Stony Brook University, Stony Brook, New York 11794, USA
[4]Department of Physics, The Ohio State University, Columbus, Ohio 43210, USA
[5]Johannes Gutenberg-Universität, Institut für Physik, D-55099 Mainz, Germany
[6] Department of Chemistry, Princeton University, Princeton, New Jersey 08544, USA

## I. Computational Details

We performed density functional theory (DFT) calculations using the Quantum ESPRESSO code [1] and one shot GW and GW plus Bethe Salpeter equation (GW-BSE) calculations with the BerkeleyGW package [2]. Calculations were performed in a supercell configuration with 30 Å of vacuum in the aperiodic $z$ direction. Monolayers of $WS_2$ and $MoSe_2$ were relaxed with the Perdew-Burke-Ernzerhof (PBE) generalized gradient approximation (GGA) [3], using a 350 Ry wave function cutoff and a $24 \times 24 \times 1$ uniform k-grid, obtaining lattice constants of 3.187 Å and 3.322 Å respectively for $WS_2$ and $MoSe_2$. The Mo 4s and 4p, Se 3d, and W 5s and 5p orbitals are treated as semicore states in the pseudopotential.

For the GW and GW-BSE calculations, we generate DFT wavefunctions in a planewave basis with a kinetic energy cutoff 100 Ry, and sample the full Brillouin Zone (BZ) on a uniform $12 \times 12 \times 1$ k-grid. To overcome the slow convergence of the dielectric matrix in 2D with k-point sampling [4,5], we employ the nonuniform neck subsampling (NNS) method [5] to perform a nonuniform k-point sampling equivalent to a $2286 \times 2286$ k-grid. The Coulomb interaction is truncated [6] in the z-direction to cut off non-physical long-range Coulomb interaction between periodic images. In the GW calculation, we employ a 35 Ry cutoff energy for the planewave components of the dielectric matrix. Frequency-dependence is included through the Hybertsen-Louie Generalized Plasmon Pole (HL-GPP) [7] model. The semicore states in the pseudopotential are excluded from the charge density in the calculation of the GW self-energy.

Regarding the convergence with the summation over unoccupied bands for both the polarizability and the GW self-energy, we use 20,000 explicitly calculated DFT bands that are then compressed using the stochastic pseudobands method [8] to reduce the computational expense. After convergence test, 5 pseudobands per subspace ($N_\xi$) and 400 stochastic subspaces ($N_S$) are sufficient to generate pseudobands that introduce less than 10 meV error in the calculation of the quasiparticle (QP) energy. All valence bands and 10 conduction bands are kept exactly without further stochastic sampling. The convergence of band edges and gaps with the summation over unoccupied bands are shown in Fig. S1 (a) and (b) respectively.

---

[*] diana.qiu@yale.edu
[*] thomas.allison@stonybrook.edu

For the BSE calculations, we use $24 \times 24 \times 1$ coarse grids and interpolate them to $300 \times 300 \times 1$ fine grids to converge the absorption spectra and exciton energies. Since the band-edge excitons are composed of transitions in the vicinity of the K point, to reduce computational cost, we restrict the BSE calculations to a uniformly sampled patch of 0.193 Å$^{-1}$ around the K point in BZ [4,9].

Finally, for the supercell calculations of the 43.9° twisted heterostructure, we generate the wavefunction using plane waves with a kinetic energy cutoff of 60 Ry and sample the entire BZ on a $3 \times 3 \times 1$ k-grid using the NNS method to achieve an effective sampling of $571 \times 571$ k-grid. A 25 Ry cutoff energy is used for the dielectric matrix and HL-GPP model is used to account for frequency dependence in the screening.

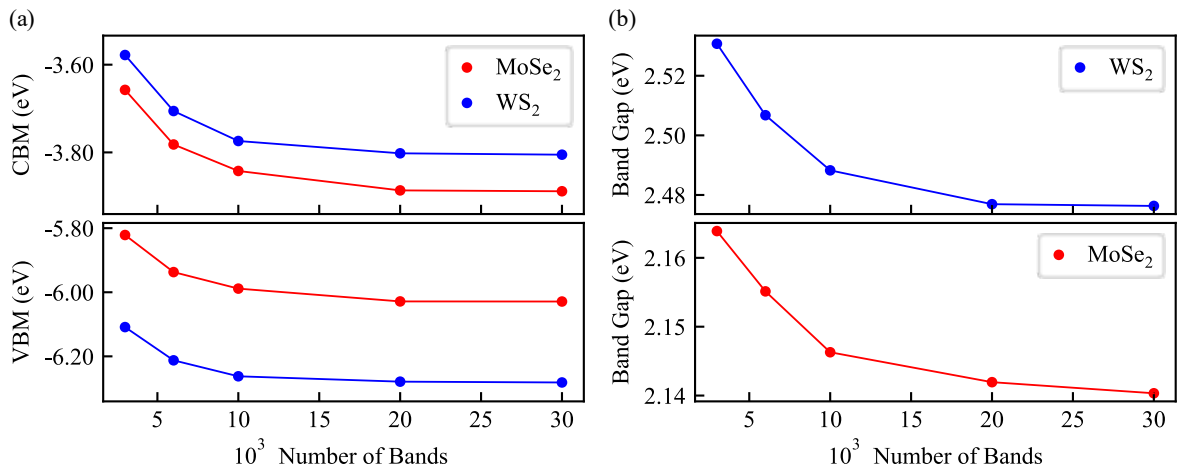

Fig. S1 (a) QP conduction band minimum (CBM) and valence band maximum (VBM) at the K point in the BZ as a function of the number of bands summed. The error in the energy becomes smaller than 10 meV once the number of bands exceeds 20,000, indicating that 20,000 bands are sufficient for achieving converged calculations. (b) QP band gaps of MoSe$_2$ and WS$_2$. In comparison to the convergence behavior of absolute band energies shown in (a), the convergence of band gaps is significantly faster, with 6,000 bands being sufficient to converge the band gap within 40 meV.

**II. Impact of Strain, Monolayer Screening, and Bilayer Screening on Band Energies**

In the main text, we performed GW calculations on MoSe$_2$ and WS$_2$ with bilayer screening and observed that the band alignment of the MoSe$_2$/WS$_2$ heterostructure changes under varying strain. The shifts in absolute energy and band gap under strain are consistent with previous studies [10,11]. However, these calculations do not identify the most critical factor influencing the band alignment: strain, monolayer screening, or bilayer screening. To address this, we performed additional calculations at both the DFT and GW levels with monolayer screening, as shown in Fig. S2, alongside the bilayer-screening GW calculations presented in the main text.

At the DFT level, only strain affects the band energy, and no GW correction is applied. These results indicate that strain alone is sufficient to alter the band alignment. For GW calculations with both monolayer and bilayer screening, while the band gap is significantly affected, the band alignments in $H_X^M$ and $H_X^X$ remain unchanged. Thus, it is the local reconstruction in different high-symmetry regions that causes the coexistence of type I and type II band alignments in the MoSe$_2$/WS$_2$ heterostructure.

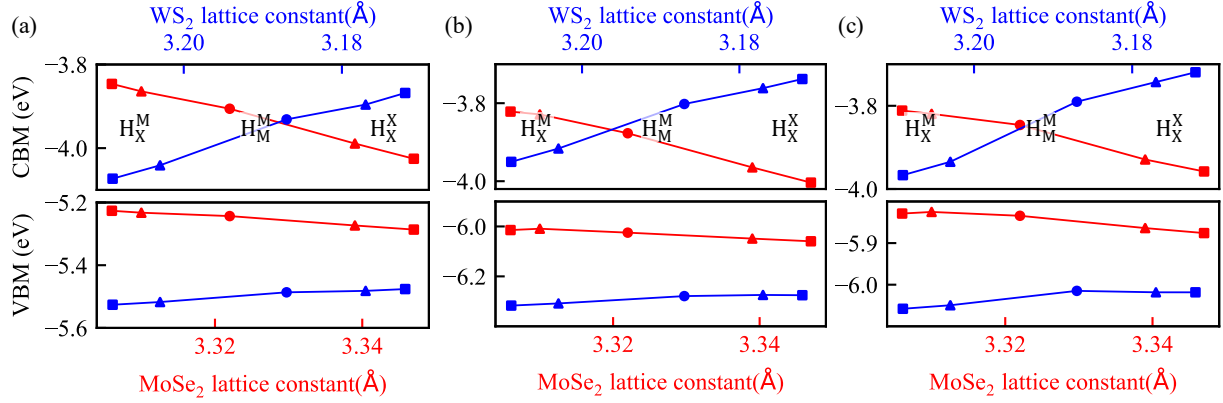

Fig. S2 Band energies as a function of lattice constants at (a) DFT level, (b) GW level with monolayer screening, and (c) GW level with bilayer screening. Circular points are unstrained lattice constant corresponding to the average lattice constant in the $H_M^M$ stacking configuration. Square points are average length of the lattice constants in high-symmetry regions $H_X^M$ and $H_X^X$. Triangular points are maximum or minimum length of the lattice constant appearing at the boundary of regions $H_X^M$ and $H_X^X$, which demonstrate that the band alignment within one high-symmetry region remains consistent.

## III. Strain Distribution at Different Twist Angles

As a supplement to Table 1 in the main text, we present the strain distribution in the supercells at various twist angles to visualize the atomic reconstruction.

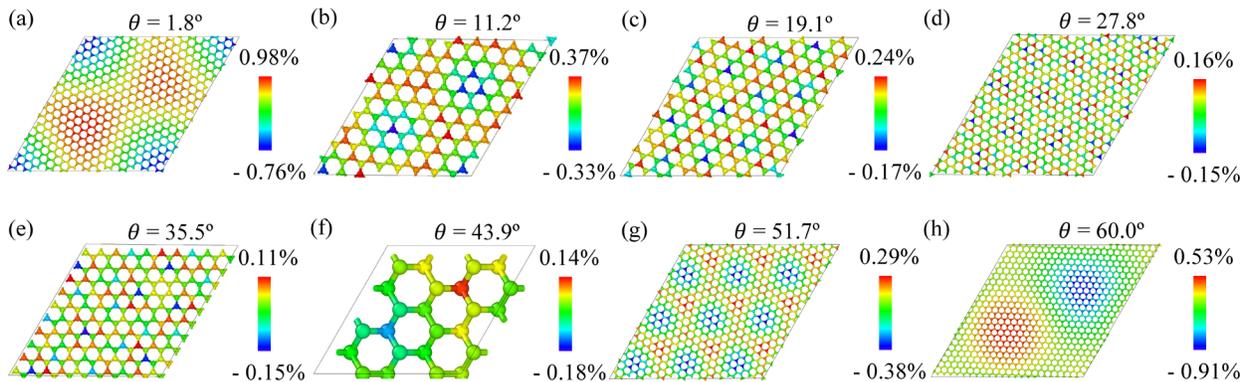

Fig. S3 (a)-(h) Strain distribution in the supercells at various twist angles, with maximum and minimum strain percentages indicated. Positive values represent compressive strain, while negative values correspond to tensile strain.

## IV. Standard Deviation of Lattice Constants in Different High-Symmetry Regions

In the main text, we treat the lattice in high-symmetry regions as uniform to facilitate the GW and GW-BSE calculations. To justify the approximation, we choose a radius of 1.2 nm, comparable to the calculated exciton radius [4], for the three regions, and calculate the standard deviation of local lattice constants, which is defined as the distance between the nearest metal atoms, to evaluate the uniformity. For each region with a radius of 1.2 nm, there are 62 metal atoms and thus 156 lattice constants. Standard deviations in different high-symmetry regions of the two layers are shown in Table S1.

Since the radius of excitons reported in experiments can be as large as 1.7 nm [12,13], we also expand the radius of high-symmetry regions to 1.6 nm for checking the robustness of our approximation, which is also shown in Table S1. We find that even for 1.6 nm radius, most of the standard deviations are still around 0.005 Å and the largest one appearing in $WS_2$ $H_M^M$ is less than 0.01 Å. Therefore, it is reasonable to use a uniform unit cell for GW and GW-BSE calculations in high-symmetry regions.

Table S1. Standard deviations in high-symmetry regions of both layers with two different radii.

| Radius (nm) | Regions | $MoSe_2$ $H_X^M$ | $MoSe_2$ $H_X^X$ | $MoSe_2$ $H_M^M$ | $WS_2$ $H_X^M$ | $WS_2$ $H_X^X$ | $WS_2$ $H_M^M$ |
|---|---|---|---|---|---|---|---|
| 1.2 | | 0.002 Å | 0.005 Å | 0.006 Å | 0.003 Å | 0.005 Å | 0.007 Å |
| 1.6 | | 0.003 Å | 0.006 Å | 0.007 Å | 0.004 Å | 0.006 Å | 0.009 Å |

## V. Sample Fabrication and Characterization

The $WS_2$/$MoSe_2$/hBN heterostructures were fabricated using mechanical exfoliation and dry transfer method. First, hexagonal boron nitride (hBN) was mechanically exfoliated onto a highly doped n-type silicon substrate (0.001–0.005 Ω·cm) using Scotch tape. An hBN flake with a thickness of 20–30 nm was selected via optical contrast to construct the heterostructure. Then, monolayer $WS_2$ and $MoSe_2$ were mechanically exfoliated onto a polydimethylsiloxane (PDMS) substrate. Second harmonic generation (SHG) spectroscopy was further performed to control the twist angle between the layers. Finally, the monolayer $MoSe_2$ and $WS_2$ were sequentially transferred onto the hBN flake using the dry transfer method. Some part of the exfoliated $MoSe_2$ or $WS_2$ flake was draped over the edge of the hBN to ensure a conductive pathway exists and sample charging does not occur during ARPES measurements.

The heterostructures with $MoSe_2$ on $WS_2$ have twist angles 58.04°±0.31° and 40.19°±0.87°. Photoluminescence and reflection contrast measurements from the monolayer and heterobilayer regions are shown in Fig. S4 and S5 along with the optical and PEEM images of the sample. These optical measurements were acquired at 90 K. The heterostructures with $WS_2$ on $MoSe_2$ have twist angles 59.1°±1.4° and 40.0°±0.4°. The photoluminescence from monolayer and heterobilayer regions, the optical image, and the PEEM image are shown in Fig. S6 and S7. These measurements were acquired at room temperature.

## VI. Time-Resolved ARPES Measurements and Data Analysis

The Stony Brook tr-ARPES beamline and its application to 2D materials has been described previously [14–17]. All measurements were done with vacuum pressure at or below $9 \times 10^{-10}$ torr and were measured with 25.2 eV probe photon energy. All measurements are performed at room temperature and with a pump wavelength of 517 nm (2.4 eV). The ML $MoSe_2$, ML $WS_2$, and

40° MoSe$_2$ on WS$_2$ bilayer were excited with 101 $\mu J/cm^2$ pump fluence. The 58° MoSe$_2$ on WS$_2$ bilayer was excited with 193 $\mu J/cm^2$. The pump pulses were s-polarized during these measurements. The 59° WS$_2$ on MoSe$_2$ bilayer was excited with 43 $\mu J/cm^2$ and the 40° WS$_2$ on MoSe$_2$ bilayer was excited with 45 $\mu J/cm^2$. The pump pulses were p-polarized during these measurements. For the bilayer samples, data are collected from a ~10 μm region of interest (ROI) selected using a field aperture in a real-space image plane of the microscope [18].

The data obtained from the instrument is post processed to eliminate image distortions and for image calibration. First, the surface photovoltage (SPV) produced from the pump pulse exciting the Si substrate is corrected for and energy axes referenced to the valence band maxima (VBMs) are generated. The procedure for this correction has been described previously [16,17]. In ML-WS$_2$, the VBM is determined by fitting an EDC of the spin-orbit split bands at K with a double Gaussian and using the fitted peak position as the VBM. For ML-MoSe$_2$, the ARPES signal at the VBM at K is very weak due to the photoemission matrix element, so the VBM is determined by fitting an EDC from the Γ band with a Gaussian, then adding previously reported 380 meV [19] to determine the true VBM at K. For the bilayer samples, the signal at K is very weak in general so all energies are referenced to the local valence band maximum at Γ, $E_\Gamma$. Since the electronic states at Γ have contributions from both layers, we do not expect the position of $E_\Gamma$ observed in ARPES to depend on the layer orientation. Within the experimental error, Γ and K are energetically degenerate in the bilayer samples.

Once the energy axes are determined, each 3D image $(k_x, k_y, E)$ is normalized to the counts within a large ROI around Γ. This normalization accounts for any drift in the photoelectron yield at different pump-probe time delays. A background signal $(k_x, k_y, E)$ is constructed from the average of data taken with negative pump-probe time delays and is subtracted from the remaining signal $(k_x, k_y, E, t)$.

To produce the time traces shown in Fig. 5 (g) and (h), the exciton signal at the K-point is integrated using a 0.28 Å$^{-1}$ × 0.28 Å$^{-1}$ k$_x$-k$_y$ bin size and an energy bin tuned to capture the entire exciton signal. Then, electron distribution curves (EDCs) are extracted from the same regions of interest in the background-subtracted exciton signal, summing data taken at or after 5 ps pump-probe delay. The EDCs are fit with Gaussians and the extracted peak positions are used for comparison between samples. To produce the K-Γ-K cuts, the data is distortion-corrected within the k$_x$-k$_y$ plane using thin-spline transformations obtained from symmetry guided image registration [20]. Image shear along the E-k$_x$ and E-k$_y$ axes are corrected using affine transformations. The background subtracted data is stitched to data without background subtraction to show the exciton signal and the valence band signal together. After the shear is eliminated, the data is 6-fold symmetrized in the k$_x$-k$_y$ plane. Then, a 0.10 Å$^{-1}$ bin is used to generate the K-Γ-K cut displayed and the exciton signal is multiplied by a constant so that it can be seen. Data taken at and after 5 ps pump-probe time delay is summed to show the final state after the initial dynamics have decayed. Fig. S8 and S9 show the K-Γ-K cuts on linear and natural logarithmic scales.

The systematic error is dominated by uncertainties in the calibration function that maps time-of-flight in the momentum microscope to energy. Typically features within a spectrum are compared to other features within the same spectrum. Therefore, any systematic offset from the absolute energy would be eliminated when comparing differences within that spectrum. However, here features across different datasets where the systematic errors might differ are compared. Therefore, a cancellation of the systematic error is not expected in this case.

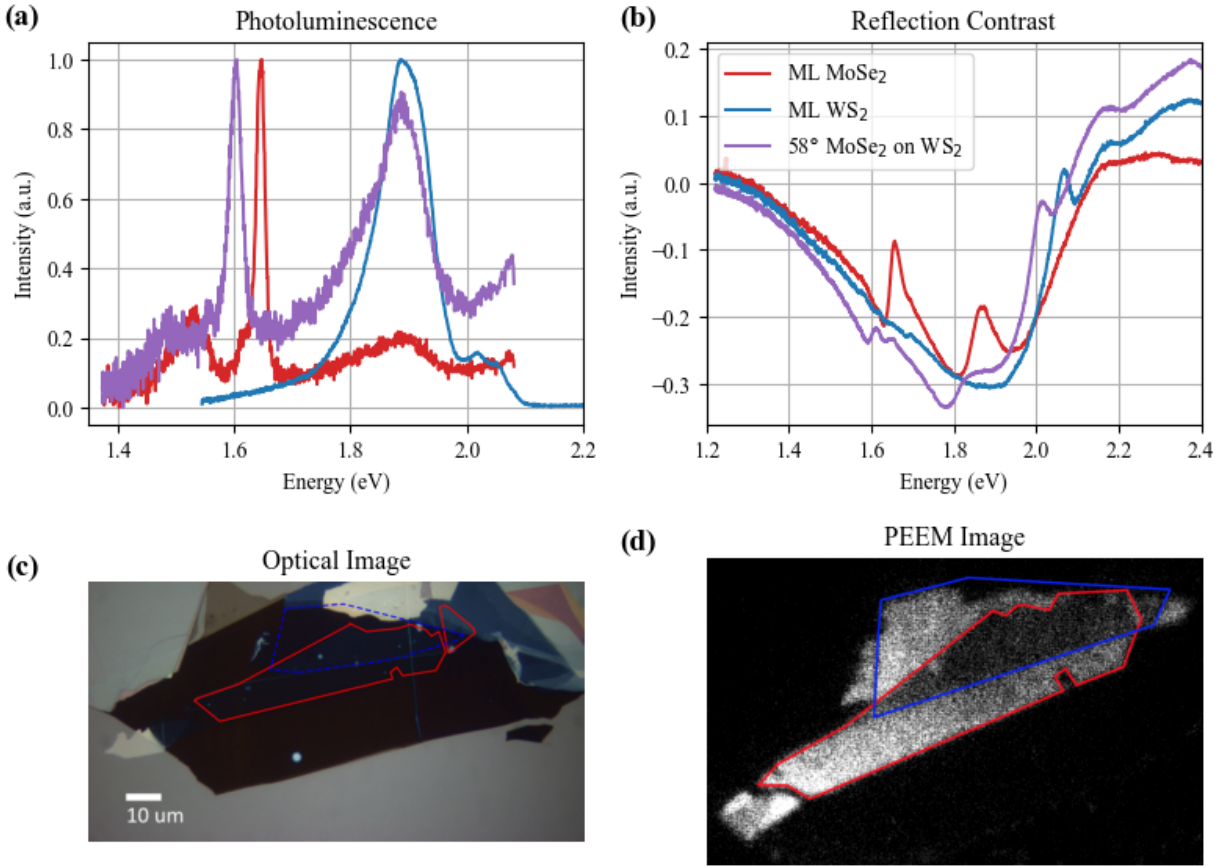

Fig S4. 58.04° ± 0.31° MoSe$_2$ on WS$_2$. (a) Photoluminescence and (b) reflection contrast from the ML MoSe$_2$, ML WS$_2$, and the bilayer regions recorded at 90 K. The legend is the same for (a) and (b). The signal near the WS$_2$ A exciton at 1.90 eV is observed in both the ML WS$_2$ and bilayer regions. In contrast, the peak near the MoSe$_2$ A exciton at 1.65 eV shifts to 1.60 eV in the bilayer. Corresponding features are observed in the reflection contrast. (c) Optical image and (d) PEEM image of the sample. The red outlines the ML MoSe$_2$ region and the blue outlines the ML WS$_2$. A 10 μm scale bar is provided in the optical image. The PEEM image was obtained with multi-photon photoemission from the 517 nm pump.

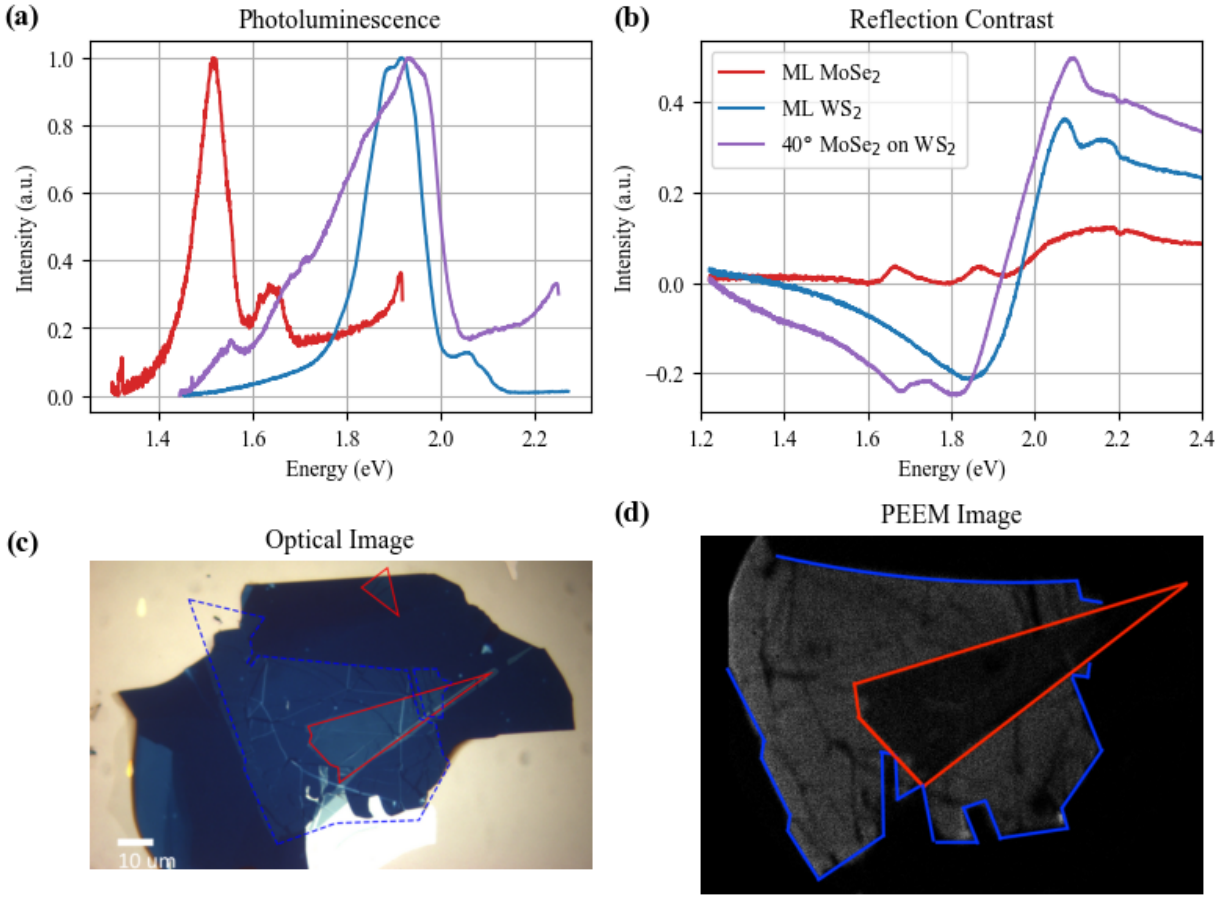

Fig S5. 40.19° ± 0.87° MoSe$_2$ on WS$_2$. (a) Photoluminescence and (b) Reflection contrast from the ML MoSe$_2$, ML WS$_2$, and the bilayer regions recorded at 90 K. The response near the WS$_2$ A exciton at 1.90 eV is observed in both the ML WS$_2$ and bilayer regions. The MoSe$_2$ PL spectrum is now dominated by the peak at 1.52 eV instead of the A exciton peak at 1.65 eV. In the bilayer PL, there is a peak at 1.55 eV. (c) Optical image and (d) PEEM image of the sample. The red outlines the ML MoSe$_2$ region and the blue outlines the ML WS$_2$. A 10 μm scale bar is provided in the optical image. The PEEM image was obtained with multi-photon photoemission from the 517 nm pump.

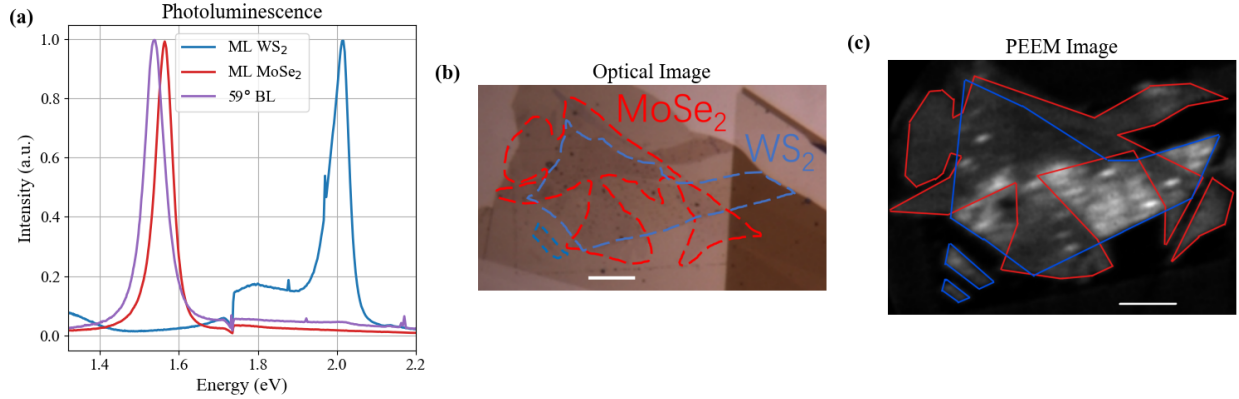

Fig S6. 59.14° ± 1.45° WS$_2$ on MoSe$_2$. (a) Photoluminescence from the ML MoSe$_2$, ML WS$_2$, and the bilayer regions recorded at room temperature. (b) Optical image and (c) PEEM image of the sample. The PEEM image was obtained with multi-photon photoemission from the 517 nm pump. The red outlines the ML MoSe$_2$ region and the blue outlines the ML WS$_2$. The white bars are approximately 20 μm.

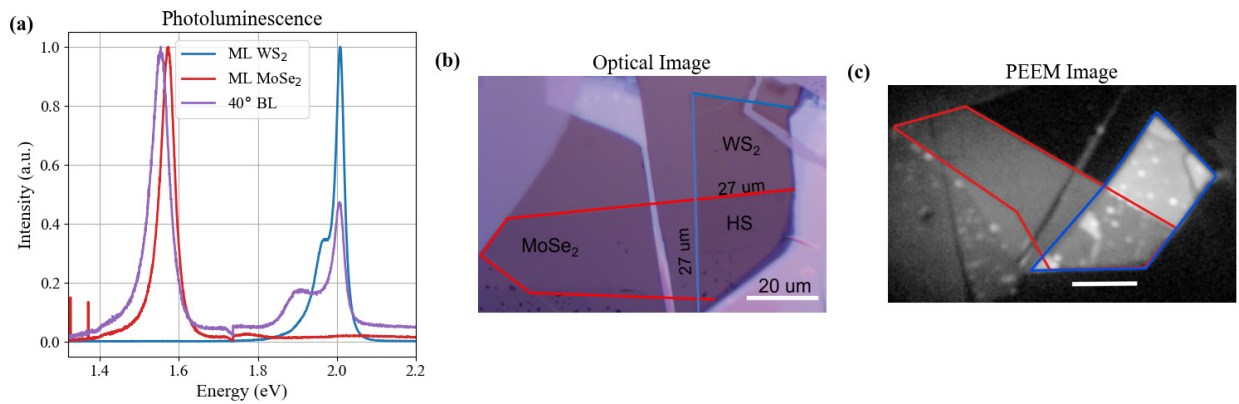

Fig S7. 40.0° ± 0.4° WS$_2$ on MoSe$_2$. (a) Photoluminescence from the ML WS$_2$, ML MoSe$_2$, and the heterostructure regions recorded at room temperature. (b) Optical image and (c) PEEM image of the sample. The PEEM image was obtained with multi-photon photoemission from the 517 nm pump. The red outlines the ML MoSe$_2$ region and the blue outlines the ML WS$_2$. The white bars are approximately 20 μm.

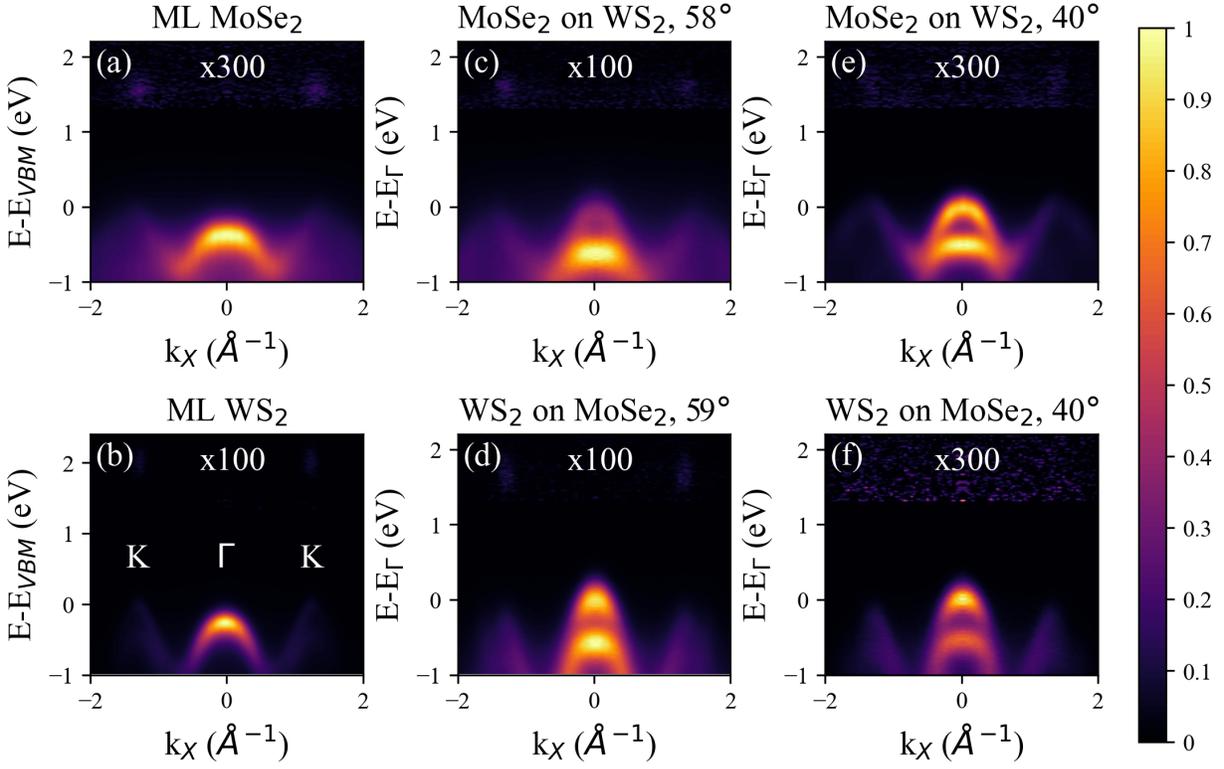

Fig S8. Same K-Γ-K cuts of ML MoSe2, ML WS2, and twisted heterobilayers as presented in Fig. 5, but on a linear scale. K-Γ-K path through the Brillouin Zone binning 0.10 Å$^{-1}$ and after 5 ps delay. The exciton signal above 1.3 eV has been background subtracted using data recorded at negative pump-probe delays and is multiplied by the constant shown on each plot. Symmetry-guided image registration [20] corrects for the aberrations and rotates the data in the $k_x$-$k_y$ plane. The shear along E-$k_x$ and E-$k_y$ is corrected using affine transformations. The data is 6-fold symmetrized and then the K-Γ-K slice is taken. The data is clipped so that the minimum value is 0, then each contour is normalized to its maximum.

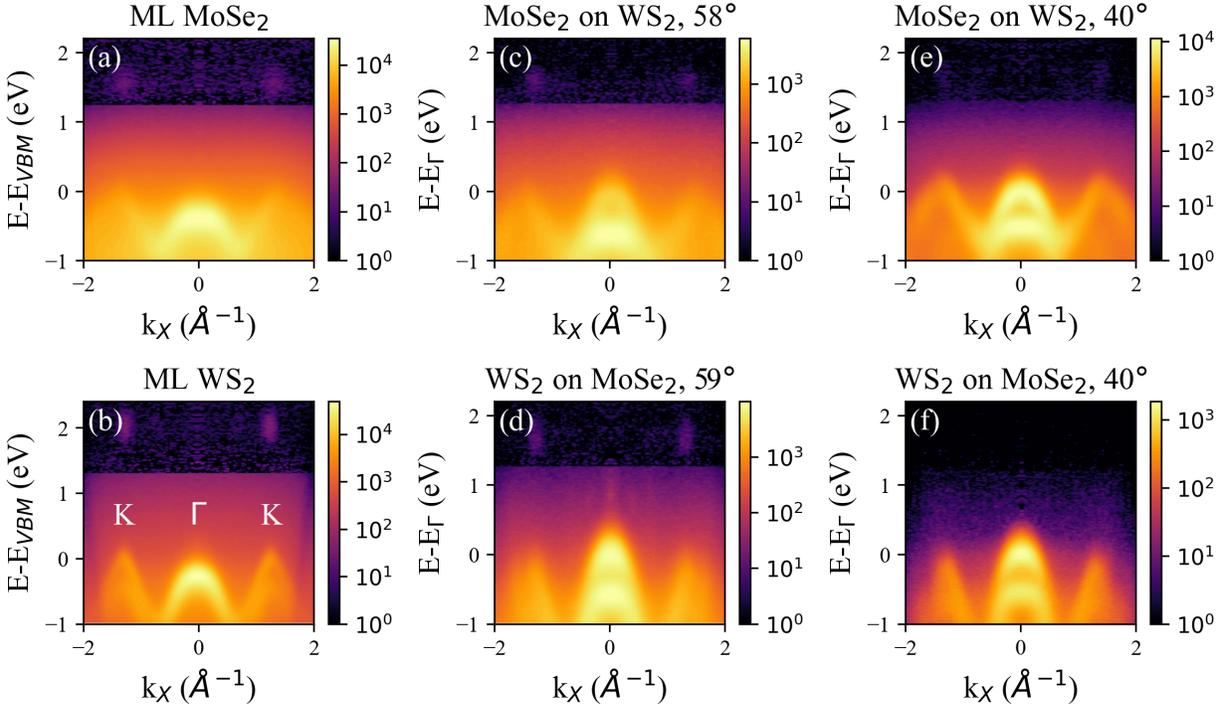

Fig S9. Same K-Γ-K cuts of ML MoSe2, ML WS2, and twisted heterobilayers as presented in Fig. 5, but on a logarithmic scale. K-Γ-K path through the Brillouin Zone binning 0.10 Å$^{-1}$ and after 5 ps delay. The exciton signal above 1.3 eV has been background subtracted. Symmetry-guided image registration [20] corrects for the aberrations and rotates the data in the $k_x$-$k_y$ plane. The shear along E-$k_x$ and E-$k_y$ is corrected using affine transformations. The data is 6-fold symmetrized and then the K-Γ-K slice is taken.